\def\dbcol{double column single spacing}
\ifdefined\dbcol
	\ifdefined\ACM
		\documentclass{sig-alternate-10pt}
		\makeatletter
		\def\@copyrightspace{\relax} 
		\makeatother
	\else
		\documentclass[conference,10pt,a4paper]{IEEEtran}
		\IEEEoverridecommandlockouts
		\makeatletter
		\def\ps@headings{%
		\def\@oddhead{\mbox{}\scriptsize\rightmark \hfil \thepage}%
		\def\@evenhead{\scriptsize\thepage \hfil \leftmark\mbox{}}%
		\def\@oddfoot{}%
		\def\@evenfoot{}}
		\makeatother
	\fi

\else 
	\documentclass[conference,11pt,onecolumn]{IEEEtran}
\fi

\usepackage[left=.7in,right=.7in,top=.7in,bottom=.6in]{geometry}

\ifdefined\ACM 
	\usepackage{amsmath}
    \usepackage{cite} 
\else
	\usepackage[cmex10]{amsmath} 
	\usepackage{amsthm} 
	\ifdefined\COMSOC 
		\usepackage[nocompress]{cite}
	\else
		\usepackage{cite} 
	\fi
	\ifCLASSINFOpdf
	  \usepackage[pdftex]{graphicx} 
	\else
	  \usepackage[dvips]{graphicx}
	\fi
\fi

\ifdefined\META     \usepackage{stmaryrd}    \fi 

\newcommand{\begproof}{\ifdefined\dbcol\begin{IEEEproof}\else\begin{proof}\fi}
\newcommand{\Endproof}{\ifdefined\dbcol\end{IEEEproof}\else\end{proof}\fi}

\ifdefined\ACM
\else
	\usepackage[font=small]{caption} 
\fi
\ifdefined\COMSOC 
  \usepackage[caption=false,font=footnotesize,labelfont=sf,textfont=sf]{subfig}
\else
  \usepackage[caption=false,font=footnotesize]{subfig}
\fi

\usepackage{amssymb}
\usepackage{bm}
\usepackage{verbatim}
\usepackage{color}
\usepackage[normalem]{ulem}
\usepackage[ruled,vlined,linesnumbered]{algorithm2e}
\usepackage{enumitem} 
\usepackage{bbm} 

\usepackage{url}

\newcommand{\fref}[1]{Fig.~\ref{#1}}

\newcommand{\cref}[1]{Corollary~\ref{#1}}

\newcommand{\aref}[1]{Algorithm~\ref{#1}}

\newcommand{\inv}[1]{\frac{1}{#1}} 

\usepackage{fancyhdr}
\fancypagestyle{empty}{
  \fancyhf{}
  \fancyhead[C]{\footnotesize{T. Luo and S. G. Nagarajan, ``Distributed anomaly detection using autoencoder neural networks in WSN for IoT'', {\it IEEE ICC}, May 2018.}}
}

\title{Distributed Anomaly Detection using Autoencoder Neural Networks in WSN for IoT}
\author{
	\IEEEauthorblockN{Tie Luo\IEEEauthorrefmark{1} and Sai G. Nagarajan\IEEEauthorrefmark{2}\\
	\IEEEauthorblockA{
	\IEEEauthorrefmark{1}Institute for Infocomm Research, A*STAR, Singapore\\
	\IEEEauthorrefmark{2}Engineering Systems and Design Pillar, Singapore University of Technology and Design\\
	E-mail: luot@i2r.a-star.edu.sg, sai\_nagarajan@mymail.sutd.edu.sg} } } 

\begin{document}
\maketitle
\thispagestyle{empty}

\begin{abstract}
Wireless sensor networks (WSN) are fundamental to the Internet of Things (IoT) by bridging the gap between the physical and the cyber worlds. Anomaly detection is a critical task in this context as it is responsible for identifying various events of interests such as equipment faults and undiscovered phenomena. However, this task is challenging because of the elusive nature of anomalies and the volatility of the ambient environments. In a resource-scarce setting like WSN, this challenge is further elevated and weakens the suitability of many existing solutions. In this paper, for the first time, we introduce autoencoder neural networks into WSN to solve the anomaly detection problem. We design a two-part algorithm that resides on sensors and the IoT cloud respectively, such that (i) anomalies can be detected at sensors in a fully distributed manner without the need for communicating with any other sensors or the cloud, and (ii) the relatively more computation-intensive learning task can be handled by the cloud with a much lower (and configurable) frequency. In addition to the minimal communication overhead, the computational load on sensors is also very low (of polynomial complexity) and readily affordable by most COTS sensors. Using a real WSN indoor testbed and sensor data collected over 4 consecutive months, we demonstrate via experiments that our proposed autoencoder-based anomaly detection mechanism achieves high detection accuracy and low false alarm rate. It is also able to adapt to unforeseeable and new changes in a non-stationary environment, thanks to the unsupervised learning feature of our chosen autoencoder neural networks.
\end{abstract}


\section{Introduction}\label{sec:intro}

Wireless sensor networks (WSN) are the eyes and ears of the Internet of Things (IoT) in the sense that they transform physical phenomena into digital signals and transfer these signals to the interconnected cyber-world for much richer processing and analytics. In reality, however, such signals or data are rarely perfect, where anomalies often arise and can interfere with the subsequent IoT analytics. {\em Anomalies}, also known as {\em outliers}, are data that do not conform to the patterns exhibited by the majority of the data set \cite{ad09survey}. It is important to identify anomalies because they often indicate events of interest such as equipment faults, sudden environmental changes, security attacks, and so on. In fact, anomaly detection is even {\em essential} to many IoT systems when, oftentimes, the whole purpose of deploying a sensor network as part of an IoT ecosystem is not to know the ``norm'' but to capture the highly erratic event occurrences.

Although the task of anomaly detection could be undertaken by a central IoT entity such as ``back-end'' as often referred to, such a scheme tends to cause highly inefficient resource utilization (besides undesirable delay). This is because of the large amount of raw data that needs to be transmitted from sensors to the central entity, which entails substantial channel interference and energy consumption, while in fact only a very small fraction of the transmitted data are anomalous.

Therefore, it is more desirable to push such tasks to the ``edge'' as much as possible, to maximize the efficiency of resource utilization as well as responsiveness. However, anomaly detection is challenging for resource-limited  sensors, owing to the elusive nature of anomalies and the volatility of ambient physical environment. Existing solutions based on various approaches have been proposed in the literature, such as threshold-based detection and Bayesian assumptions of prior distributions \cite{rajasegarar2008anomaly}, classification using k-nearest neighbors \cite{xie2013scalable}, local messaging based distributed detection\cite{chen2015distributed,branch2013network}, support vector machines (SVM) based detection\cite{zhang2013distributed}, and so forth. However, they are either computationally expensive or incur large communication overhead, which weekends their applicability to resource-constrained wireless sensor networks.

In this paper, we propose to use autoencoder neural networks \cite{vincent2010stacked} for anomaly detection in WSN. {\em Autoencoders} are a deep learning model, and are traditionally used in image recognition \cite{vincent2010stacked} and other data mining problems such as spacecraft telemetry data analysis\cite{autoenc14mlsda}. However, deep learning is generally not an option for WSN because of its formidable hunger for computational resources. We overcome this infeasibility by building an autoencoder neural network that consists of only three layers including the one hidden layer of neurons. This simple structure, however, performs very well due to the inherent power of reconstructing the input data by autoencoders. Specifically, we design a two-part algorithm that resides on sensors and the IoT cloud, respectively, such that (i) anomalies can be detected at sensors in a fully distributed manner, without the need for communicating with any other sensors or the cloud, and (ii) the relatively more computation-intensive learning task (for model training) can be handled by the cloud. The communication between sensors and the cloud happens at a much lower (and configurable) frequency than sensing. In addition, the computational load is also very low on sensors (in polynomial rather than exponential complexity such as in \cite{xie2013scalable}), which is friendly to resource-limited sensor hardware. 

For evaluation purposes, we use a sensor dataset collected over four consecutive months from a WSN testbed deployed in our office building. We demonstrate that our proposed distributed anomaly detection using autoencoders achieves high detection accuracy and low rate of false alarms (jointly characterized by a comprehensive metric called AUC). We also demonstrate that it can adapt to unforeseeable and new changes (which are however common in reality), which substantiates its suitability for non-stationary and evolving environments \cite{o2014anomaly}.

To the best of our knowledge, this work is the first to introduce autoencoders into WSN, or more broadly IoT, for anomaly detection. Besides the minimal communication and computation requirements and the distributed advantage, we also reap the benefit from the {\em unsupervised learning} feature that autoencoders possess (while most neural networks do not). This is particularly useful because it dissolves the common challenge that supervised machine learning models often lack of sufficient training data of anomalies.


\section{Related Work}\label{sec:relwork}


There are various anomaly detection methods with different levels of complexity. The basic idea is to model the distribution of what is considered to be ``normal" and then check if the target data deviate from the distribution to a significant degree. For example, several statistical techniques as noted by \cite{rajasegarar2008anomaly} assume a prior distribution for the ``normal'' data and perform a hypothesis test based on that assumption. However, model mismatch will occur when data in the real world do not adhere to the assumption. Some other techniques take a threshold-based approach as also surveyed by \cite{rajasegarar2008anomaly}, but it is difficult to identify a good threshold, and even if found, it inevitably pertains to particular setting and is hard to generalize. 

In view of such problems, non-parametric methods were proposed as more sophisticated solutions. For example, \cite{xie2013scalable} uses the k-nearest neighbors algorithm to create a hyper-grid around a given data point, and consider the data point anomalous if less than $k$ other data points lie inside that hyper-grid. This algorithm has a computational complexity of $O\left(2^{M-1}\right)$, where $M$ is the dimension of input data which can be hundreds or even thousands, making it prohibitive for high-dimensional cases. 

In the quest of searching for more efficient solutions, \cite{chen2015distributed} and \cite{branch2013network} propose distributed algorithms that rely on message exchange among sensors to detect anomalies in real time. Although these methods are computationally less intensive, they require frequent and reliable communications among sensors, which imposes serious power consumption on energy-constrained sensors. 

To reduce communication overhead, \cite{zhang2013distributed} propose two distributed online anomaly detection techniques based on a hyper-ellipsoidal one-class support vector machine (SVM). The main idea is to take advantage of the spatiotemporal correlation between sensor data and to update the SVM model (more specifically the ellipsoidal boundary) to reflect the change in the normal sensor data. However, these techniques involve matrix inversions which are computationally unfriendly to sensors. For a more comprehensive survey, the reader is referred to \cite{o2014anomaly}.

We propose to use autoencoder neural networks for undertaking the anomaly detection task in WSN. We leverage the reconstruction ability of autoencoders and design a two-part algorithm that eliminates the need for communication between sensors. The algorithm running on sensors only involves a simple matrix dot product operation whose computational complexity is much lower than existing solutions.

Autoencoders were traditionally used in image recognition problems \cite{vincent2010stacked} and spacecrafts' telemetry data analysis \cite{autoenc14mlsda}, which all have (statically) well-defined training and test data sets. In a distributed and networked context like WSN, which is much more complex and dynamic, this paper---to the best of our knowledge---is the first work that uses autoencoders for anomaly detection in a networked context.



\section{Model and Method}\label{sec:model}

\subsection{Preliminaries and Model}\label{sec:prelim}

\begin{figure}[ht]
\centering
\fbox{\includegraphics[width=0.8\linewidth]{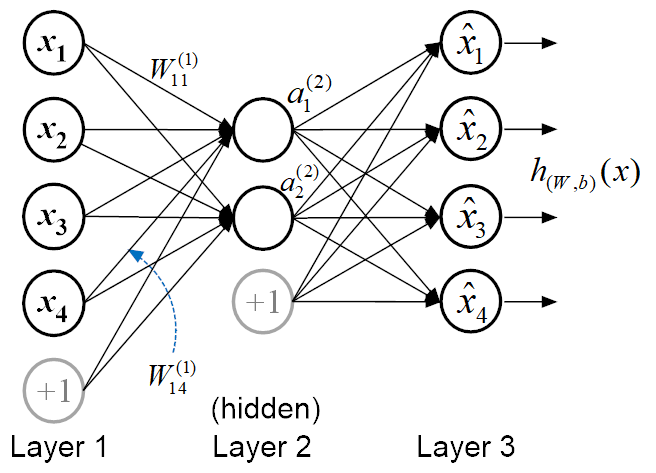}}
\caption{An autoencoder neural network.}
\label{fig:autoenc}
\end{figure}

An artificial neural network is an interconnected group of processing nodes, i.e., ``neurons'', that jointly perform a (typically nonlinear) transformation of inputs to certain desired outputs. An autoencoder is a special type of neural networks whose objective is to {\em reconstruct} the inputs instead of {\em predicting} some target variables. By reconstructing inputs, an autoencoder tries to learn a condensed representation of the input data, a process also known as ``encoding''. Formally, see \fref{fig:autoenc} in which an autoencoder consists of:
\begin{enumerate}
\item An input layer: an $M$-dimension vector that represents the input signals, denoted by $\mathbf x = (x_1,x_2,...,x_M)$. For example, it could be the pixel values of an image, or a time series of temperature sensor readings. 

\item An output layer: denoted by vector $\mathbf{\hat x} = (\hat x_1,\hat x_2,...,\hat x_M)$. Note that this is distinct from general neural networks where we would use $\mathbf y = (y_1,y_2,...,y_{|\mathbf y|})$ to denote the output layer. In the case of autoencoders, the output has the same dimension as the input, and we would like  the output to be {\em equal} to the input for the purpose of {\em reconstructing} the original input. Hence we automatically obtain our training samples by setting $\mathbf y = \mathbf x$, which is why autoencoders are unsupervised learning models. 

\item One or multiple hidden layers: sitting between the input and output layers, these hidden layers aim to learn a pattern in the inputs so as to ``encode'' the essential information (with affordable information loss). 
Let us denote the total number of all the layers by $L$ and index the layers by $l=1,2,...,L$, and denote the number of nodes in the $l$-th layer by $n^{(l)}$ (which does not count the {\em bias unit} ``+1'' which we explain shortly). For example in \fref{fig:arch}, $L=3$ and $n^{(2)}=2$.

In general, an autoencoder has multiple hidden layers, but in this particular work we use the simplest form---one layer only---to minimize the computational overhead for distributed detection. It was generally observed to perform reasonably well in practice, as also demonstrated by our own evaluation.

\item Activation function and hyper-parameters: each neuron 
in layer $l=2,3,..$ represents an {\em activation function} $f(\cdot)$ 
which is typically a sigmoid function:\footnote{The other common choice for $f$ is the hyperbolic tangent, or {\tt tanh}, function. Recently, a rectified linear activation function was discovered as yet another choice which often works well for deep learning.}
\[ f(z) = \frac{1}{1 - e^{-z}}. \]

The (hyper-)parameters of an autoencoder are weights $\mathbf W$ and biases $\mathbf b$, where $W_{ij}^{(l)}$ is the weight associated with the connection from node $j$ in layer $l$ to layer $i$ in layer $l+1$,\footnote{The reversed order of subscript indexes might look counter-intuitive but this notational convention was adopted by the neural networks literature likely because of {\em back-propagation}.} and $b_i^{(l)}$ is the bias associated with node $i$ in layer $l+1$. For example, if we denote by $a_i^{(l)}$ the output value (a.k.a. {\em activation}) of the neuron $i$ in layer $l=2,3,...$, then
\begin{align*}
a_1^{(2)} &= f (W_{11}^{(1)} x_1 + W_{12}^{(1)} x_2 + ... + W_{14}^{(1)} x_4 + b_1^{(1)} ) \\
a_2^{(2)} &= f (W_{21}^{(1)} x_1 + W_{22}^{(1)} x_2 + ... + W_{24}^{(1)} x_4 + b_2^{(1)} )
\end{align*}
In a more compact form, and more generally for all $l=2,3,...$, we write
\begin{align}\label{eq:io_basic}
\mathbf a^{(l)} = f (\mathbf W^{(l-1)} \mathbf a^{(l-1)} + \mathbf b^{(l-1)} ).
\end{align}
The intercept term $\mathbf b^{(l-1)}$ in the above corresponds to the ``+1'' node (in $n^{(l)}$ copies) at layer $l-1$, which is called a {\em bias unit} or bias node.

The final output, $\mathbf{\hat x}$, is then $\mathbf a^{(L)}$ which we also denote by $h_{(W,b)}(\mathbf x)$.
\end{enumerate}

Given $T$ training samples $\{\mathbf x[1],\mathbf x[2],...,\mathbf x[T]\}$ where each $\mathbf x[i]=(x_1[i],x_2[i],...,x_M[i])$, the objective of an autoencoder is to minimize the cost function
\begin{align}
\begin{split}
J(W,b) := &\inv T \sum_{i=1}^T \left( \inv 2 || h_{W,b}(\mathbf x[i]) - \mathbf x[i] ||^2 \right) \\
&+\frac{\lambda}{2} \sum_{l=1}^{L-1} \sum_{j=1}^{n^{(l)}} \sum_{i=1}^{n^{(l+1)}} \left( W_{ij}^{(l)} \right)^2
\end{split}
\end{align}
The first term defines the {\em reconstruction error} with respect to the original inputs, and the second term is a regularization term to prevent overfitting. Minimizing this function can be solved using standard gradient descent methods such as stochastic gradient descent or L-BFGS (Limited-memory Broyden–Fletcher–Goldfarb–Shanno) algorithms \cite{haykin2009neural,ufldl-tut}.

\subsection{System Architecture and Two-Part Algorithm}

\begin{figure}[t]
\centering
\fbox{\includegraphics[width=\linewidth]{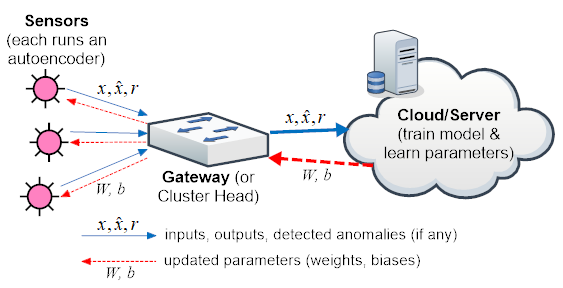}}
\caption{Architecture of a WSN that uses autoencoders for anomaly detection.}
\label{fig:arch}
\end{figure}

Our system architecture is presented in \fref{fig:arch}. Each sensor runs one copy of the autoencoder and performs two tasks (apart from sensing): (i) provide the input and output of the autoencoder to the IoT cloud as the training data, which are uploaded via a gateway or cluster head in a much lower frequency than sensing (e.g., once a day versus once every two minutes), 
(ii) perform anomaly detection, which is done locally without the need for communicating with any other sensors, the gateway, or the IoT cloud. The gateway simply relays data between sensors and the cloud, and can be bypassed if each sensor is equipped with an Internet connection capability such as 3G, WiFi or Ethernet. 

The cloud trains the autoencoder machine learning model using the training data ($\mathbf x, \mathbf{\hat x}$) provided by {\em all} the sensors. Afterward, it sends the updated model parameters ($\mathbf W,\mathbf b$) back to the sensors to update their respective autoencoders.\footnote{This describes the case when the sensors are monitoring the same physical phenomenon in proximity, where at a given point in time, sensor readings are similar across sensors if no anomaly occurs. For a large-scale WSN, 
the sensors can be organized into $c$ clusters (hence the ``cluster head'' in \fref{fig:arch}) such that each cluster is represented by an autoencoder and the cloud will simply maintain $c$ autoencoders.} In addition, if there are anomalies reported by the sensors, the cloud may take actions such as alarming to a control center or conducting IoT analytics such as root cause analysis.

Specifically, each sensor performs anomaly detection as follows. For the ease of description, we assume the frequency of uploading training data to be once a day. Hence, if the sensing frequency is once every two minutes, then the dimension of the input vector is $M=720$. Note that it is well known that one aggregated transmission is much more energy-efficient than multiple separate transmissions even if the data size is larger; in fact, one can reduce the communication overhead even further by using a (lossless) local data compression algorithm such as S-LZW~\cite{s-lzw06} or LEC~\cite{LEC09}.
\begin{enumerate}
\item Each sensor, say $s$, calculates the reconstruction error, or {\em residual}, for the $m$-th input-output observation ($m=1,2,...,M$) on a given day $d$:
\[ r_m(s,d) = x_m(s,d) - \hat x_m(s,d). \]

\item Sensor $s$ uploads $\{r_m(s,d)|m=1,2,...,M\}$ to the cloud at the end of the day $d$.

\item The cloud calculates the statistics of the residual in terms of mean and variance:
\begin{align}\label{eq:errdist}
\begin{split}
\mu_m &= \inv{DS} \sum_{d=1}^D \sum_{s=1}^S r_m(s,d) \\
\sigma_m^2 &= \inv{DS} \sum_{d=1}^D \sum_{s=1}^S \left( r_m(s,d) - \mu_m \right)^2
\end{split}
\end{align}
where $D$ is the total number of days over which the cloud computes the statistics, and $S$ is the total number of sensors.

\item \label{step:newstat} The cloud sends $\mu_m$ and $\sigma_m$ back to every sensor.

\item \label{step:detect} Each sensor detects anomaly by calculating
\begin{align}\label{eq:detect}
\alpha_m(s,d) = \begin{cases}
0, \text{ if } |r_m(s,d) - \mu_m| \le p \sigma_m \\
1, \text{ otherwise}
\end{cases}
\end{align}
where $p$ is a parameter chosen according to the specific use case. Typically, the residuals $r$ are Gaussian and we can choose $p=2$ (or $p=3$) which corresponds to that 5\% (or 2.5\%) observations on the average would be classified as anomalies.

Note that step \ref{step:detect} does not need to wait for step \ref{step:newstat} which simply allows sensors to update their {\em current} values of $\mu_m$ and $\sigma_m$.
\end{enumerate}

This is essentially a two-part algorithm that resides in part on sensors and in part on the cloud. Thus, the pseudo-code of the above is presented in \aref{alg:dada-s} (DADA-S) and \aref{alg:dada-c} (DADA-C). Prior to executing the algorithms, we initialize the parameters $\mathbf W$ and $\mathbf b$ as well as $\bm\mu$ and $\bm\sigma$ by training our autoencoder model over a historical data set with no or negligible anomalies (this training phase also involves presetting the ``seed'' values of $\mathbf W$ and $\mathbf b$ which we set as small random numbers as suggested by \cite{ufldl-tut}).

\begin{algorithm}
\caption{DADA-S: Distributed Anomaly Detection using Autoencoders (Sensor's algorithm)}\label{alg:dada-s}
\For{$d \gets 1$ \KwTo $\infty$} {
Obtain sensor readings $\mathbf x = \{x_1,x_2,...,x_M\}$\;\label{line:sense}
Feed $\mathbf x$ into autoencoder to obtain output $\mathbf{\hat x} = (\hat x_1,\hat x_2,...,\hat x_M)$\; \label{line:autoenc_io}
Calculate residual vector $\mathbf r = \mathbf x - \mathbf{\hat x}$\;
Detect anomaly according to \eqref{eq:detect} and obtain $\bm\alpha(s,d)$\;\label{line:detect}
Send $\mathbf x,\mathbf{\hat x},\mathbf r(s,d),\bm\alpha(s,d)$ to cloud (via gateway)\;
Update $\mathbf W,\mathbf b,\bm\mu,\bm\sigma$ received from cloud\;
}
\end{algorithm}

\begin{algorithm}
\caption{DADA-C: Distributed Anomaly Detection using Autoencoders (Cloud's algorithm)}\label{alg:dada-c}
\For{$d \gets 1$ \KwTo $\infty$} {
Receive $\mathbf x, \mathbf{\hat x}, \mathbf r(s,d), \bm\alpha(s,d)$ from all the sensors $s=1,2,...,S$\;
Store $\mathbf x, \mathbf{\hat x}$ in the training data set and react to $\bm\alpha(s,d)$ if needed\;
\If{$d$ mod $D_u=0$} {
	\tcp{\small $D_u$ denotes training frequency (in no. of days) chosen by cloud}
	Retrain autoencoder with the updated training data set; \tcp{\small data will be pre-shuffled to avoid learning biases toward the latest data}
	Recalculate $\bm\mu,\bm\sigma$ using $\mathbf r(s,d), \bm\alpha(s,d)$ according to \eqref{eq:errdist}\;
	Send updated $\mathbf W,\mathbf b,\bm\mu,\bm\sigma$ to all the sensors\;
}
}
\end{algorithm}

For maximal clarity, the algorithms have assumed a day to be the computational period. For mission-critical applications where anomalies need to be detected in real-time, it can be achieved by simply modifying the DADA-S algorithm such that lines \ref{line:sense}--\ref{line:detect} are executed when each single reading $x_m$ is obtained.\footnote{In a highly erratic environment and for more accurate result, one may want to increase the uploading frequency so that the cloud can re-train he model using a corresponding weight sub-matrix and bias sub-array, as the order of the inputs is preserved.} Note that this does not necessarily increase communication overhead as the sensor only needs to send $\alpha_m(s,d)$ when it is equal to 1 (which only happens sporadically).


The computational complexity at sensors is dominated by Line \ref{line:autoenc_io} which computes the output of the autoencoder given an input vector. In fact, this only involves a simple matrix {\em dot product} operation (cf. \eqref{eq:io_basic}) with a maximum dimension of $M$, which has a computation time of $O(M^2)$. This is much lower than $O(2^{M-1})$ as in \cite{xie2013scalable}, and is readily affordable by most COTS sensors.

\section{Performance Evaluation}

\subsection{WSN Testbed and Dataset}
\begin{figure}[ht]
\centering
\subfloat[Sensor at a corridor.]{\fbox{\includegraphics[width=0.5\linewidth]{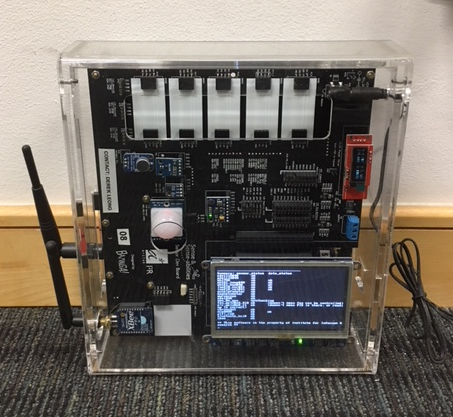}}}\hfil
\subfloat[Sensor at cubicle 1]{\fbox{\includegraphics[width=0.5\linewidth,height=3.9cm]{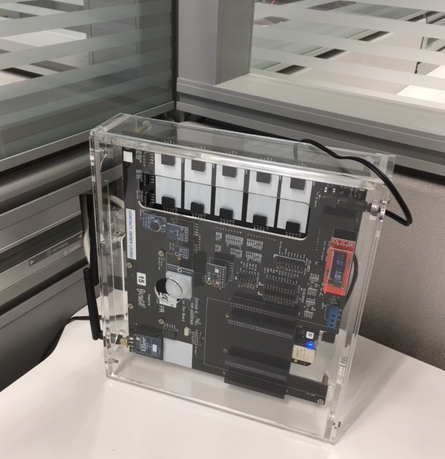}}}\vfil
\subfloat[Sensor at cubicle 2.]{\fbox{\includegraphics[width=0.5\linewidth]{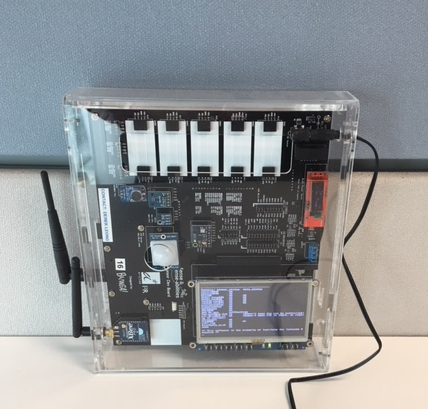}}}
\subfloat[Sensor beside a window.]{\fbox{\includegraphics[width=0.5\linewidth,height=4.1cm]{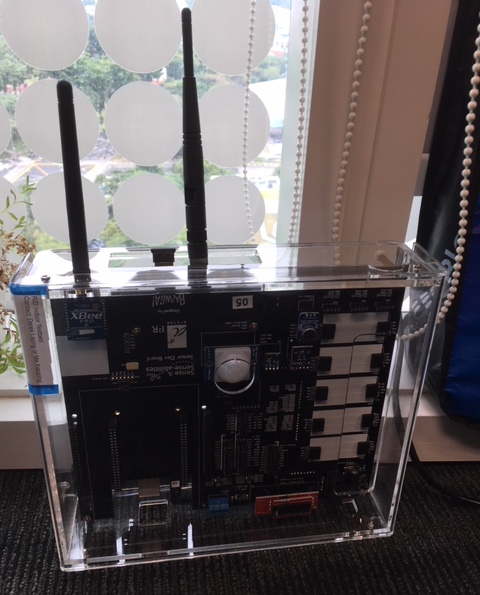}}}
\caption{The indoor deployment of our wireless sensor network.}
\label{fig:deployment}
\end{figure}

We evaluate our proposed distributed anomaly detection algorithm using an indoor WSN testbed deployed in our office building at different locations (\fref{fig:deployment}). This testbed consists of $S=8$ sensor nodes that monitor temperature and relative humidity, with a sensing frequency of once every 2 minutes, or 720 daily readings from a single sensor. We have collected all the sensor data from September to December 2016.

Since it is difficult to have sufficient real anomalies, we generate synthetic anomalies into the data set using two commonly used models \cite{fault09fusion}:
\begin{itemize}
\item {\em Spike}: a sharp rise followed immediately by a sharp decline in the sensor reading. Formally,
\begin{align}
x'(t) =  x(t) + v \delta(t),
\end{align}
where $\delta(t)$ is the Dirac delta function and $v$ is the magnitude of the spike.

\item {\em Burst}: a continuous and constant offset persisting for a finite period of time. Formally,
\begin{align}
x'(t) = \begin{cases} x(t) + v, & t_{start} \le t \le t_{end}\\
x(t), &\text{otherwise.} \end{cases}
\end{align}
\end{itemize}
In both cases, $v$ can be negative.

\subsection{Experimental setup}

We build our autoencoder neural network with the input and output layers both having $n^{(1)}=M=720$ nodes, and determine the number of hidden neurons, i.e., $n^{(2)}$, using $k$-fold cross validation. The optimal value is $n^{(2)}=504$ which corresponds to a compression ratio of 30\% ($1-504/720$). We initialize the parameters ($\mathbf W,\mathbf b,\bm\mu,\bm\sigma$) using a small portion of the dataset obtained from our WSN testbed without anomalies. In the evaluation of our algorithm, we characterize the anomaly detection performance using Area Under the Curve (AUC) corresponding to the Receiver Operating Characteristic (ROC) curves \cite{roc04notes}. An ROC curve is most commonly used to visualize the performance of a binary classifier, in terms of true positive rate (TPR, or probability of detection) and false positive rate (FPR, or false alarm rate), and AUC is arguably the best way to summarize the classifier performance in a single number. Intuitively, AUC is the probability that a classifier assigns a higher score to a randomly chosen positive-class object than to a randomly chosen negative-class object. A good classifier has an AUC close to 1 and a bad one close to 0; an AUC of $0.5$ corresponds to a random guessing classifier.

\subsection{Results}

\begin{figure}[ht]
\centering
\fbox{\includegraphics[trim=8mm 5mm 16mm 12mm,clip,width=0.495\linewidth]{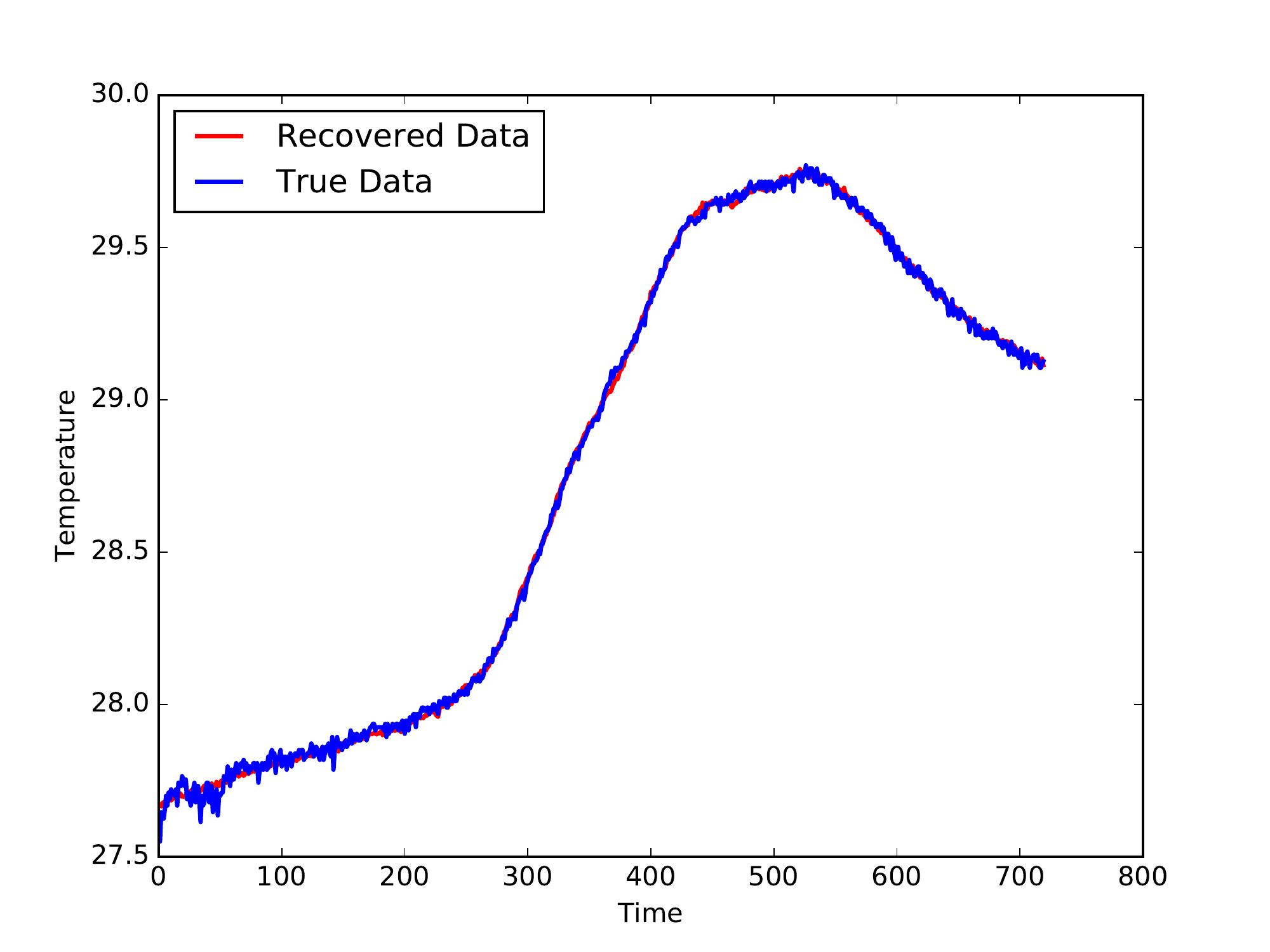}}\hfil
\fbox{\includegraphics[trim=8mm 5mm 16mm 12mm,clip,width=0.495\linewidth]{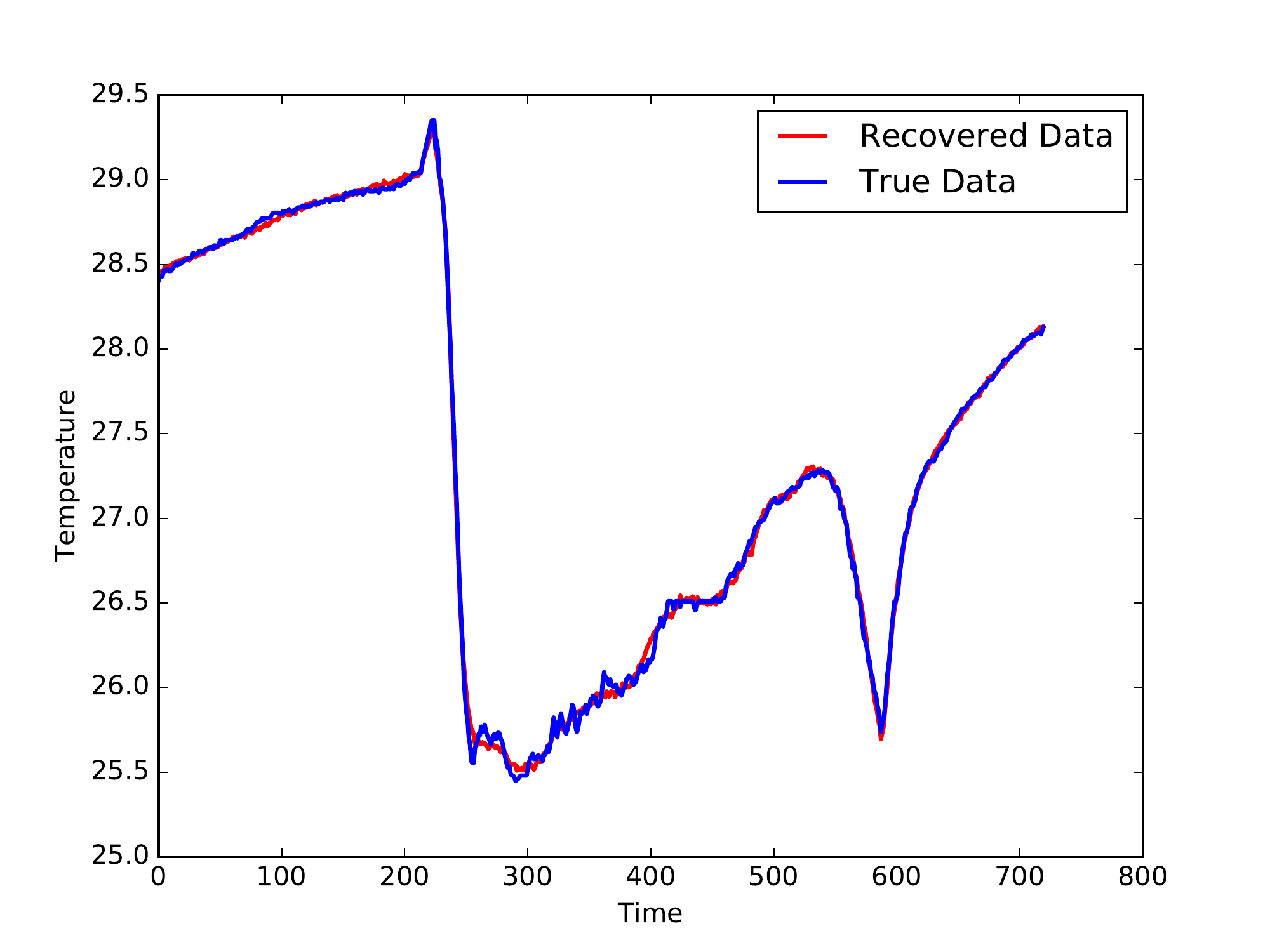}}
\caption{Reconstruction performance of the autoencoder we built. The recovered data ($\hat x$) almost coincides with the true data ($x$), indicating a valid autoencoder.}
\label{fig:reconstruct}
\end{figure}

{\bf Pre-Validation:}
We first verify if the autoencoder model has been trained properly, by looking at its reconstruction performance when there is no anomaly present. The results are shown in \fref{fig:reconstruct} for two different sensors on two different days (other sensors on other days demonstrated the similar performance). We can see that the reconstructed (or recovered) data $\hat x$ almost coincides with the original input $x$ (true data), which validates our model to proceed to the next step, anomaly detection.

\begin{figure}[ht]
\centering
\fbox{\includegraphics[trim=10mm 4mm 26mm 17mm,clip,width=0.9\linewidth]{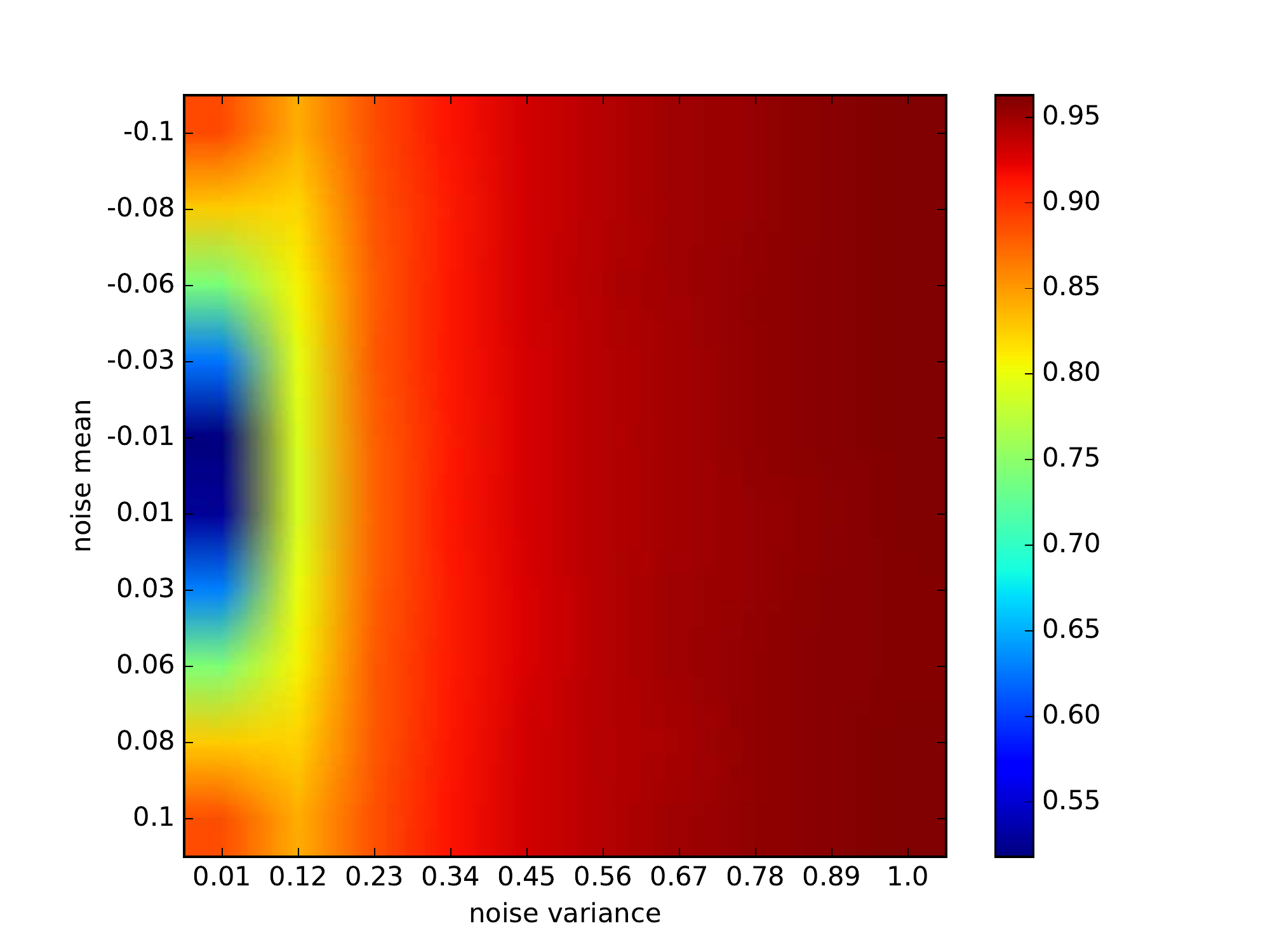}}
\caption{AUC heat map: AUC versus $\mu_v$ (Y-axis) and $\sigma_v^2$ (X-axis).}
\label{fig:auc_map}
\end{figure}

{\bf Varying anomaly magnitude:}
In this set of experiments, we vary the magnitude $v$ of spikes and bursts such that $v$ of each anomaly follows a normal distribution $\mathcal N(\mu_v, \sigma_v^2)$, while fixing the frequency of anomalies at $K=100$ per day. To see how AUC is affected by two parameters $\mu_v$ and $\sigma_v^2$, we plot AUC as a heat map against both $\mu_v$ and $\sigma_v^2$ as shown in \fref{fig:auc_map}. We see that AUC $>0.8$ most of the time, which indicates a good classifier. The exception when AUC is lower (between 0.5 and 0.8) happens when both $|\mu_v|$ and $\sigma_v^2$ are very small ($|\mu_v|<0.07$ and $\sigma_v^2<0.12$). This is because in those cases, the anomalies are not really notable, or perhaps even not anomalies. Since in practice, only {\em significant} deviations are typically concerned with, our algorithm is suitable for all such use cases. Moreover, \fref{fig:auc_map} also shows that AUC is symmetric about the mean $\mu_v$, which is well understood since positive and negative deviations are treated the same by our neural network model (captured by residual $r$).

\begin{figure}[ht]
\centering
\subfloat[$\mu_v=0.1$, $\sigma_v^2=0.01$.]{\label{fig:auc_mean0.1}
\fbox{\includegraphics[trim=8mm 4mm 16mm 14mm,clip,width=0.5\linewidth]{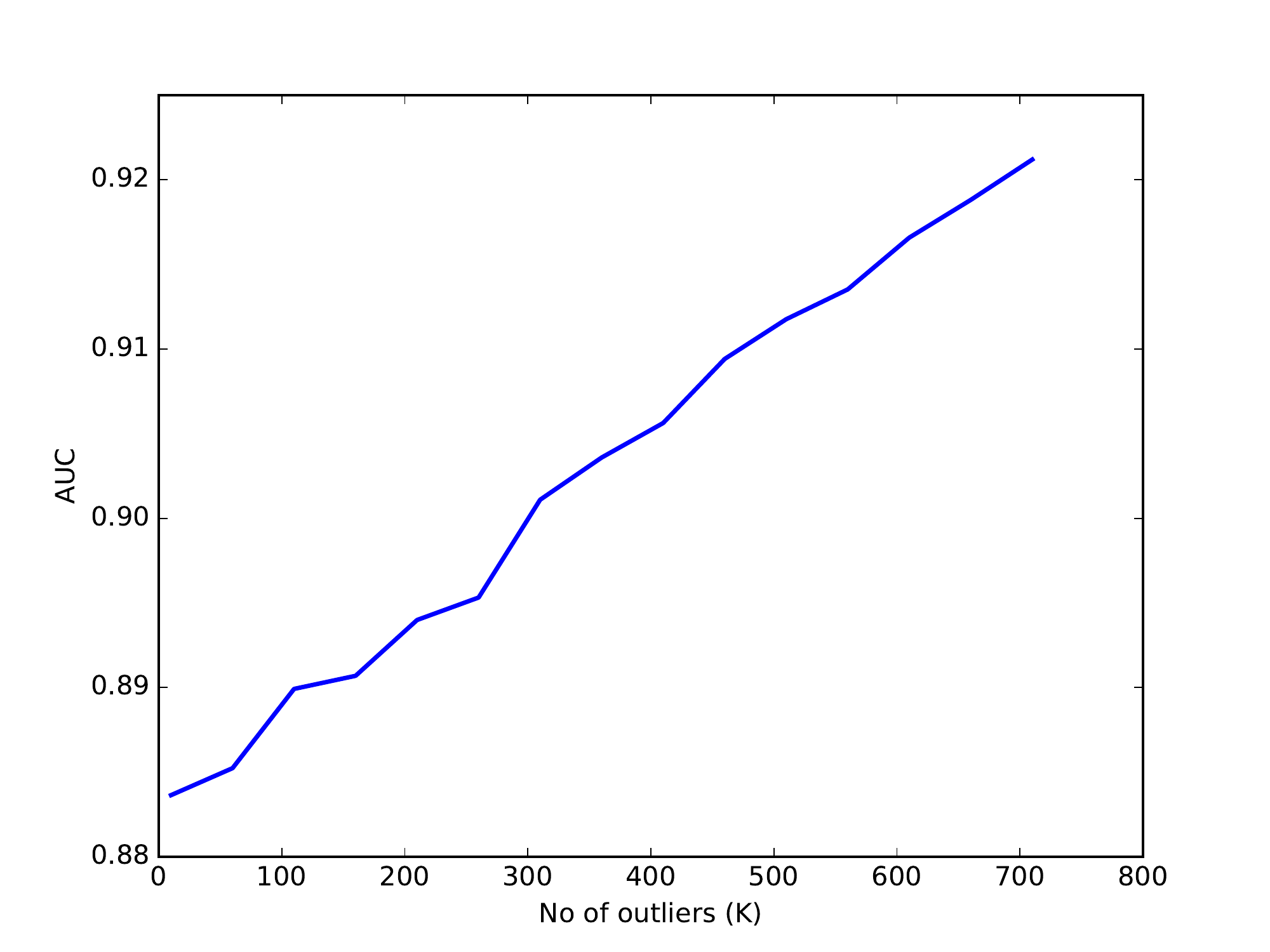}}}
\subfloat[$\mu_v=0.01$, $\sigma_v^2=0.01$.]{\label{fig:auc_mean0.01}
\fbox{\includegraphics[trim=6mm 4mm 16mm 13mm,clip,width=0.5\linewidth]{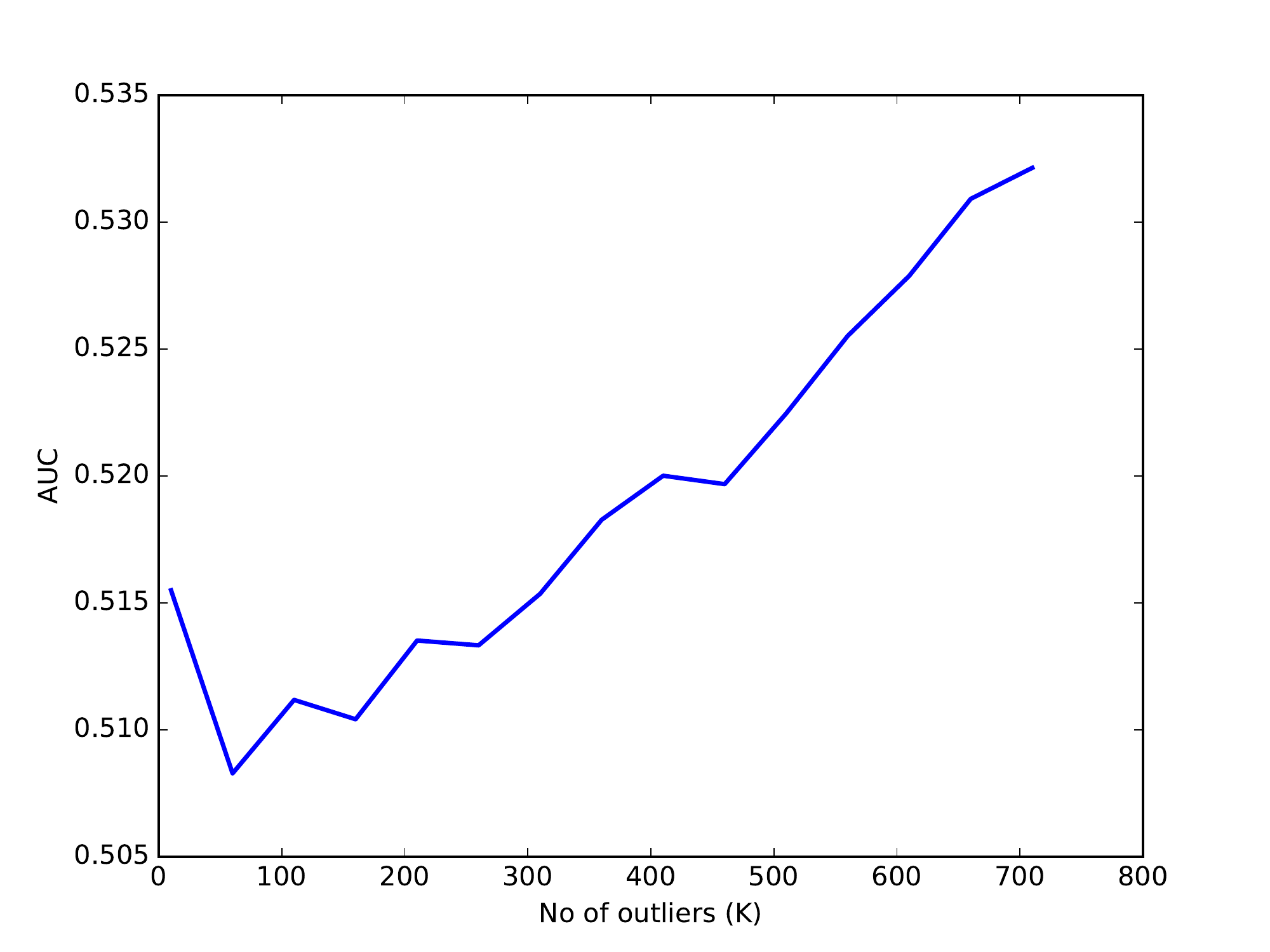}}}
\caption{AUC versus the anomaly frequency $K$.}
\label{fig:auc_vs_k}
\end{figure}

{\bf Varying anomaly frequency:}
In this set of experiments, we vary the frequency $K$, the number of anomalies per day, and present the results in \fref{fig:auc_vs_k} with the corresponding $|\mu_v|$ and $\sigma_v^2$ values. 
In general, when more and more anomalies occur, detecting anomalies becomes harder because the dominance of ``normal'' data which set the ``baseline'' of a classifier, is jeopardized. However, as we can see from \fref{fig:auc_vs_k}, the AUC of our autoencoder continues to increase (slightly). This is because AUC is not a singular metric like misclassification error but a compound metric that summarizes both TPR and FPR. Therefore, we would rather be not too optimistic but draw a more conservative conclusion from this set of results, that the performance can be maintained within a reasonable range of $K$. This reflects certain {\em robustness} of our autoencoder.

One may wonder why the algorithm performs well even when $K=720$, which corresponds to that every sensor reading is an anomaly. This is because there are 8 sensors in our testbed and these $K$ anomalies are injected into the whole data set per day. In other words, even at $K=720$, there are still 7 normal sensors on the average which help to screen out the anomalous sensor readings.

{\bf Adaptivity to non-stationary environment:}
In reality, environment is constantly evolving; observations that were previously considered anomalies could later become ``normal'', and vice versa. Thus, it is desirable to have an {\em adaptive} detector which can learn the changes and continues to work well despite unforeseeable and dynamic changes in the underlying physical phenomenon.

To this end, we configure our autoencoder with two different setups to run DADA-C:
\begin{itemize}
\item {\em Random}: Same as the above, new observations $(\mathbf x,\mathbf{\hat x})$ are randomized together with the entire historical data set which is then fed into the autoencoder to learn hyper-parameters $(\mathbf W,\mathbf b)$.
\item {\em Prioritized}: The most recent $D_u$ days' data are mixed with anther randomly chosen $D_u$ days of historical data to form the training data set which is then fed into the autoencoder. We set $D_u=14$ for this evaluation.
\end{itemize}

\begin{figure}[ht]
\centering
\subfloat[True positive rate (TPR).]{\label{fig:adapt_TPR}
\fbox{\includegraphics[trim=3.9cm 8.5cm 4.4cm 9cm,clip,width=0.5\linewidth]{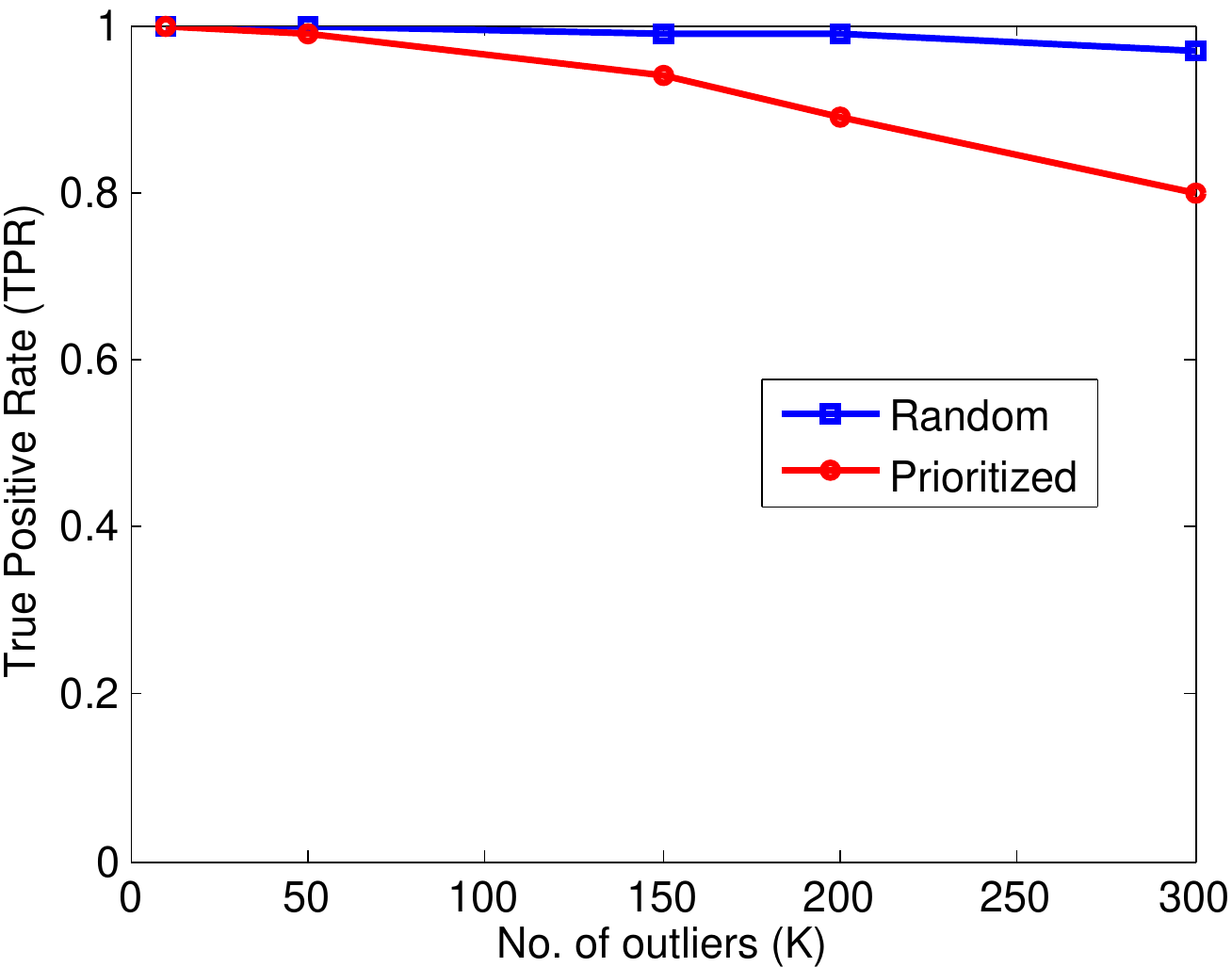}}}
\subfloat[False positive rate (FPR).]{\label{fig:adapt_FPR}
\fbox{\includegraphics[trim=3.9cm 8.5cm 4.4cm 9cm,clip,width=0.495\linewidth]{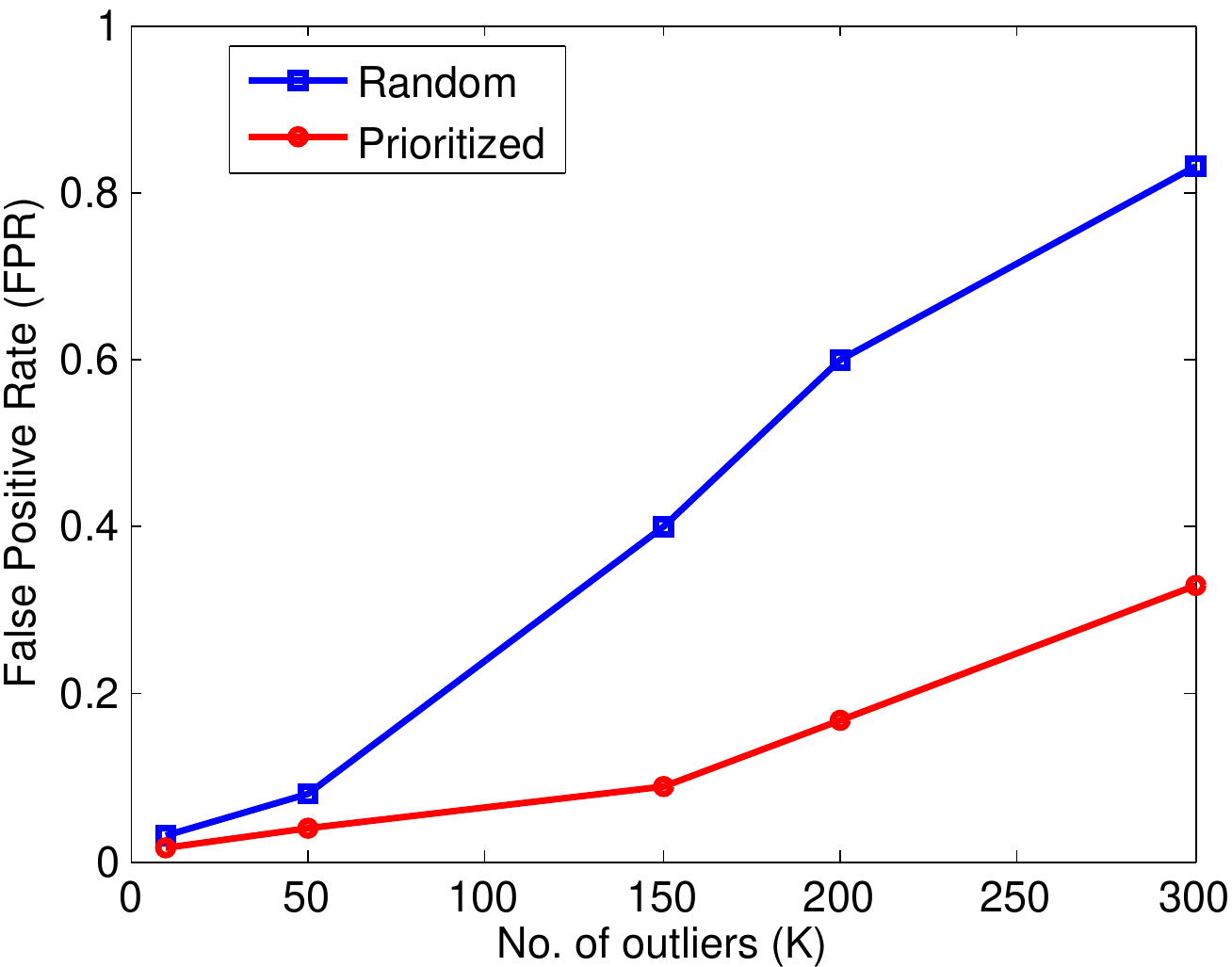}}}
\caption{Adapting to non-stationary environment.}
\label{fig:adapt}
\end{figure}

For a clearer comparison, we evaluate true positive rate (TPR) and false positive rate (FPR) separately and show results in \fref{fig:adapt}. In the evaluation, $\mu_v=0.5$ and $\sigma_v^2=0.02$, and those $v>\mu_v$ are considered true anomalies while $v\le \mu_v$ are considered acceptable environmental changes. We see that TPR slightly drops when $K$ increases. This is because the training data set is getting more and more ``perturbed'' and becomes ``less normal'', which makes anomalies less obvious to identify. The {\em Random} scheme performs better (by up to 18\%) because the majority of the training data is still historical data which is less affected by the new changes. On the other hand, in \fref{fig:adapt_FPR} we see that the {\em Prioritized} scheme has much lower false positive rate (FPR) than {\em Random}, by up to 60\%. This is because the {\em Prioritized} scheme enables the autoencoder to learn from more fresh inputs so as to update weights and biases in a more responsive manner, while as the same time reserving sufficient historical data so as to keep a balance. More specifically, the variance $\sigma_m^2$ of the residual as calculated by \eqref{eq:errdist} increases after each re-training, and hence allows for more room for the changes to be accepted without raising false alarms. 

While the most suitable choice will depend on whether the specific application is more concerned with TPR or FPR, we could recommend {\em Prioritized} in general because it strikes a more balanced tradeoff. However, the key takeaway from this set of evaluation, is that our autoencoder model is able to adapt to unforeseeable and new changes in a non-stationary environment.

\section{Conclusion}

This paper presents the first effort of introducing autoencoder neural networks into WSN to perform anomaly detection. Despite the unsuitability of deep learning in general for resource-constrained WSN, we make this approach feasible by building a simple structure of autoencoder and exploiting its powerful reconstruct-ability. The approach is made possible also by our design of a two-part algorithm specifically for spatially distributed WSN such that communication overhead is minimized and computational load is allocated to the most suitable entities. 

Specifically, our autoencoder contains a single hidden layer of neurons and the corresponding computational complexity is only polynomial ($O(M^2)$) in order to suit resource-limited sensors. Training the model involves an IoT cloud while the communication between the cloud and sensors happens at a much lower (and configurable) frequency. More importantly, anomaly detection at sensors is fully distributed and requires zero communication among sensors.

Using sensor data collected from an indoor WSN testbed over a 4-month period, we demonstrate via experiments that our proposed algorithm can detect anomalies with high accuracy and low false alarms which are jointly characterized by AUC. Furthermore, the unsupervised learning nature, as well as the flexibility in our training configuration, allows our algorithm to be able to adapt to unforeseeable and new changes in a non-stationary environment, as demonstrated in our experiments too.

In future work, we plan to extend our model to large-scale sensor networks to match the need of large-scale IoT applications seamlessly.

\bibliographystyle{IEEEtran}
\bibliography{IEEEabrv,DAD_NN}

\end{document}